\documentclass[conference]{IEEEtran}
\IEEEoverridecommandlockouts

\usepackage{amsthm}
\usepackage{amsfonts,amssymb,amsthm}
\usepackage{bbm}
\usepackage{mathtools}
\usepackage{physics}
\usepackage{amsfonts}
\usepackage[ruled,vlined,linesnumbered]{algorithm2e}
\usepackage{enumerate}
\usepackage{mathtools}
\usepackage{stackengine}

\usepackage{mathrsfs}
\usepackage{caption}
\usepackage{subcaption}
\usepackage{float}

\usepackage{tikz,pgfplots}
\usetikzlibrary{positioning,fit}

\theoremstyle{definition}
\newtheorem{definition}{Definition}

\usepackage{authblk}
\usepackage{xcolor}

\title{Finite-State Semi-Markov Channels \\for Nanopore Sequencing}
\author[$\dag$]{Brendon McBain}
\author[$\dag$]{Emanuele Viterbo}
\author[$\dag$]{James Saunderson}
\affil[$\dag$]{Department of Electrical \& Computer Systems Engineering, Monash University, Clayton, Australia}

\begin{document}

\maketitle

\begin{abstract}
    Nanopore sequencing is an emerging DNA sequencing technology that has been proposed for use in DNA storage systems. We propose the {\em noisy nanopore channel} model for nanopore sequencing. This model captures duplications, inter-symbol interference, and noisy measurements by concatenating an i.i.d. duplication channel with a finite-state semi-Markov channel. Compared to previous models, this channel models the dominant distortions of the nanopore while remaining tractable.
    Anticipating future coding schemes, we derive MAP detection algorithms and estimate achievable rates. Given that finite-state semi-Markov channels are a subclass of channels with memory, we conjecture that the achievable rate of the noisy nanopore channel can be optimised using a variation of the generalised Blahut-Arimoto algorithm.
\end{abstract}

\section{Introduction}

Nanopore sequencing \cite{ONT} has emerged as a promising technology for the sequencing of {\em deoxyribonucleic acid (DNA)}, however it is limited by many distortions imposed by the physics of the sequencer. Thus far, a complete model of the nanopore sequencer has not been established in the research community, hence the purpose of this paper is to introduce a model that is suitable for the future development of codes that enable reliable data storage in synthetic DNA when using nanopore sequencing for the reading operation \cite{Lopez2019}.

The four molecules adenine ($A$), cytosine ($C$), guanine ($G$) and thymine ($T$), connected to a sugar-phosphate backbone molecule, form the {\em nucleotides} (or {\em bases}), which are the primary elements of {\em single-stranded DNA (ssDNA)}. Abstractly, we can assume $\{A,T,C,G\}$ is the alphabet forming arbitrary length sequences. 

\begin{figure}
\includegraphics[width=\linewidth]{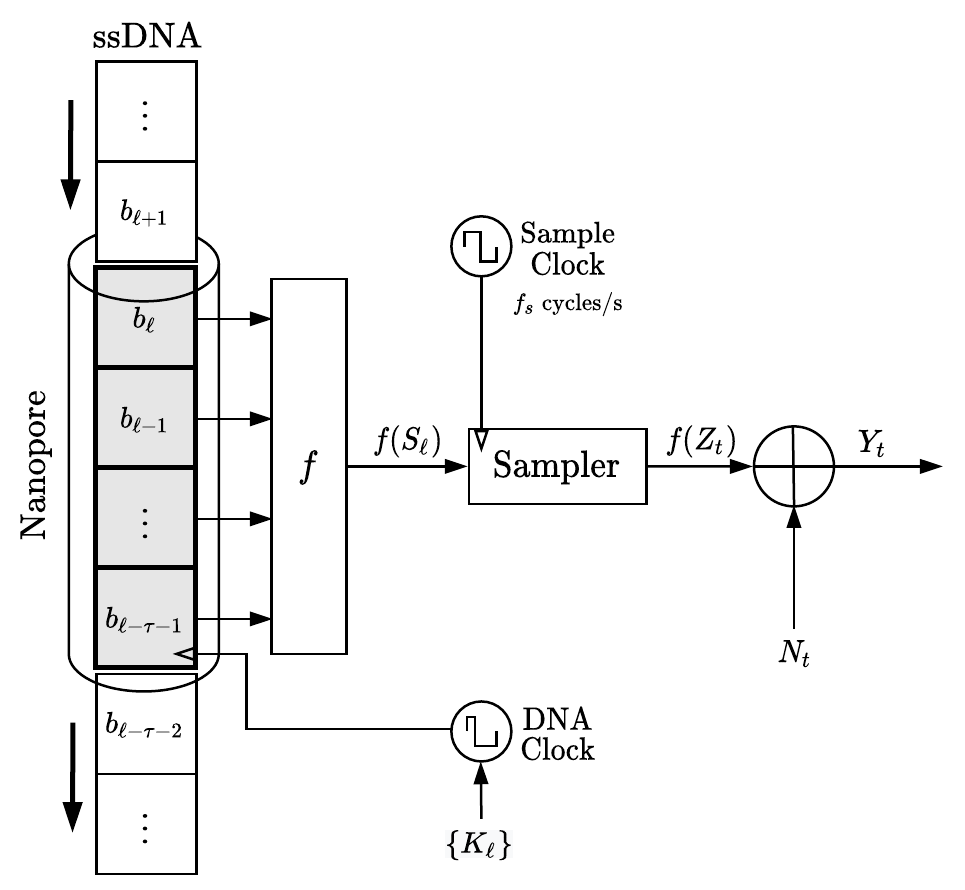}
\caption{A diagram of the noisy nanopore channel of Definition \ref{def:nnc}. The $\ell$-th base in the ssDNA is fed through the nanopore after $K_\ell/f_s$ seconds, according to the variable ``DNA Clock''. Simultaneously, the sampler samples every $1/f_s$ seconds, according to the ``Sample Clock'' with frequency $f_s$.}
\label{fig:nnc}
\end{figure}

The nanopore is a microscopic pore that holds $\tau$ nucleotides (see Fig. \ref{fig:nnc}) for a random duration determined by a motor protein that shifts nucleotides one by one through the nanopore. Meanwhile, an ionic electrical current flowing through the nanopore is uniquely disturbed by the $\tau$ nucleotides inside it at any given time, and is sampled by the sequencer at a frequency $f_s$. Measuring the current level through an output function $f$, the nanopore sequencer estimates the {\em state} of the system defined by the $\tau$ bases in the nanopore. Unfortunately, the following distortions reduce the reliability of detection \cite{Laszlo2014}:

\begin{itemize}
  \item {\bf Random dwell times (sample duplications)}: Fluctuations in the motor protein operation results in a random number of samples per nucleotide, since the nucleotide is not being shifted in lockstep with the sampler.

  \item  {\bf Fading}: For each nucleotide, the motor is designed to halt for a duration that facilitates the sampling of a constant current level. However, minor variability in the physical dimensions of the nanopores slightly changes this level. This is similar to a variable fading coefficient depending on the nanopore in use.
  
  \item  {\bf Inter-symbol interference (ISI)}: Given that the electrical current is disturbed by $\tau$ bases, the detection of a single base is distorted by inter-symbol interference with its neighbours.
  
  \item  {\bf Noisy measurements}: Finally, each sample is distorted by measurement noise.
\end{itemize}

From experimental data, the authors of \cite{Laszlo2014} additionally observed {\em backsteps (tandem duplications)} and, less frequently, {\em skippings (deletions)}.
%(attributed to the motor protein or post-processing)

Before information-theoretic methods for the nanopore sequencer can be employed, a tractable channel model that best represents the aforementioned distortions must be chosen. In this regard, the channel proposed in \cite{Mao2017} is the closest to ours that we have found in the literature, given that it modelled ISI, fading, and deletions. This channel excluded duplications by assuming a perfect segmentation algorithm, since any segmentation errors could be combined with backsteps (tandem-duplications) or deletions, but unfortunately their channel did not include backsteps to justify this decision.

Simpler channel models have been considered to study distortions in isolation, including a {\em noiseless nanopore channel} \cite{Hulett2021}, a {\em noisy (tandem-)duplication channel} \cite{Tang2019}, and a {\em memoryless nanopore channel} \cite{Conde2018}. While these channels highlight interesting properties of the nanopore, they are incomplete.

In contrast, our approach is based on a {\em finite-state Markov channel (FSMC)}  \cite[Chapter~4.6]{Gallager1968} to model ISI and noisy measurements, which we extend to include duplications. The duplications can be modelled by an i.i.d. duplication channel, 
but since sample duplications of a Markov chain result in a semi-Markov chain (or renewal process), we concatenate it with a {\em finite-state semi-Markov channel (FSSMC)} to form the proposed {\em noisy nanopore channel (NNC)}. We do not include fading, since its effect can be modelled as additional measurement noise, neither do we include deletions, since they can be eliminated by tuning  the sampling rate relative to the speed of the motor protein. 

The primary contribution of this work is establishing a tractable channel that models the dominant distortions in nanopore sequencing. Secondary contributions include generalising MAP detection algorithms that have been important tools for coding schemes of finite-state channels. In particular, we generalise the forward-backward algorithm for MAP symbol detection, and the Viterbi algorithm for MAP sequence detection. Aside from detection, the generalised forward-backward algorithm is used to estimate achievable rates using analogous methods to \cite{Kavcic2001, Arnold2001}, which leads us to conjecture that the {\em generalised Blahut-Arimoto algorithm} (GBAA) \cite{Kavcic2001} optimises the Markov source (and, if the conjecture in \cite{Kavcic2001,Vontobel2008} is true, is capacity-achieving).

This paper is organised as follows. Section II presents preliminaries on (hidden) Markov chains and (hidden) semi-Markov chains. Section III introduces the noisy nanopore channel. Section IV derives the forward-backward algorithm for MAP symbol detection, and a Viterbi algorithm for MAP sequence detection. Section V shows how to compute achievable information rates, and highlights its connection with the generalised Blahut-Arimoto algorithm. Finally, Section VI presents numerical results of computed achievable rates over different channel parameters, using a channel mapping derived from the Scrappie simulator \cite{Oxford2013}.

\section{Preliminaries}%Note that $P(K)$ is short-hand for $P(K=k)$ for all $k\in\mathsf{supp}(K)$.
\subsection{Hidden Markov chains}
A discrete-time stochastic process $(S_\ell)$ with initial probabilities $\mu_0(i) := \mathbb{P}(S_0 = i)$ is said to be a {\em Markov chain} if its {\em  Markov kernel} $P$ satisfies
\begin{align}
P(i,j) &:= \mathbb{P}(S_{\ell+1}=j|S_0,\ldots,S_\ell=i)\notag\\
&= \mathbb{P}(S_{\ell+1}=j|S_\ell=i)
\end{align}
for all $i,j\in \mathsf{supp}(S)$ and for all $\ell \in \mathbb{N}$. 
The chain $(S_\ell,Y_\ell)$ is said to be a {\em hidden Markov chain} if $(S_\ell)$ is a Markov chain and, for all $\ell \in\mathbb{N}$, it satisfies 
\begin{align}
\mathbb{P}(Y_{\ell}|S_0,\ldots, S_\ell) &= \mathbb{P}(Y_{\ell}|S_\ell)
\end{align}

\subsection{Hidden semi-Markov chains}
The discrete-time stochastic process $(S_\ell,T_\ell)$ is said to be a {\em Markov renewal process} with jump times $(T_\ell)$ if, using the parameterisation $K_{\ell+1}=T_{\ell+1}-T_\ell$, its {\em semi-Markov kernel} satisfies \cite[Definition~3.2]{Barbu2008}
\begin{multline}
    Q(i,j,k) :=\\ \mathbb{P}\bigl(S_{\ell+1}=j,K_{\ell+1}=k|S_0,\ldots,S_{\ell}=i; K_1,\ldots,K_{\ell}\bigr)\\
    = \mathbb{P}\bigl(S_{\ell+1}=j,K_{\ell+1}=k|S_{\ell}=i\bigr)
\end{multline}
for all $\ell\in\mathbb{N}$, for all $i,j\in\mathsf{supp}(S)$, and for all $k\in\mathsf{supp}(K)$. The {\em semi-Markov chain} $(Z_t)$ represents this Markov renewal chain as $Z_t = S_\ell \quad \text{if} \quad T_{\ell-1} < t \leq T_{\ell}$. We note that the main difference between chains $(S_\ell,K_\ell)$ and $(Z_t)$ is that the former is a tuple of states and durations, whereas the latter is the actual process that has a defined state at any given time, but they are equivalent. We will switch between them whenever it is convenient to do so throughout this work.

A stochastic process $(Z_t,Y_t)$ is said to be a {\em hidden semi-Markov chain} if it satisfies \cite[Definition~6.1]{Barbu2008}
\begin{align}
    \gamma_{t'+1}^{t}(i,j) &:= \mathbb{P}\left(Y_{t'+1}^{t}, Z_{t'+1}^{t}=j |Y_{1}^{t'}; Z_1,\ldots,Z_{t'}=i\right) \notag\\
    &=Q(i,j,t-t') \prod_{r=t'+1}^{t} \mathbb{P}(Y_r|Z_r = j) 
\end{align}
for all $i,j \in \mathsf{supp}(Z)$ and for all $t',t \in \mathsf{supp}(T)$ such that $i \neq j$ and $t' < t$. Assume the processes $(Y_t)$ and $(K_{\ell})$ are i.i.d. throughout this work. 

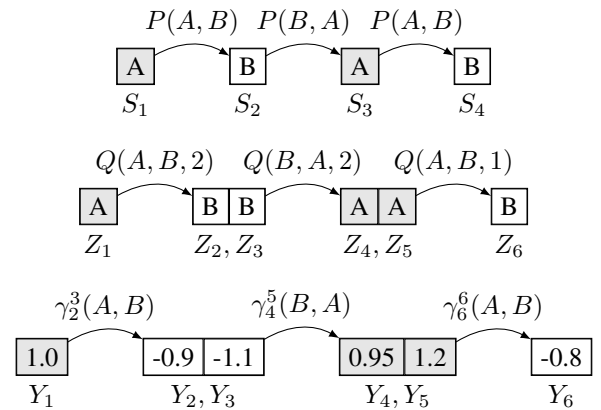
\begin{figure}[!h]
\centering

\begin{tikzpicture}
\tikzset{odd/.style={draw,rectangle,fill=gray!20}}
\tikzset{even/.style={draw,rectangle}}

\node[odd] (a1) {A};
\node[even, right=1cm of a1] (a2) {B};
\node[odd, right=1cm of a2] (a3) {A};
\node[even, right=1cm of a3] (a4) {B};

\node (af1)  [draw, inner sep=0pt,
             label=below:{$S_1$},
             fit=(a1)] {};
\node (af2)  [draw, inner sep=0pt,
             label=below:{$S_2$},
             fit=(a2)] {};
\node (af3)  [draw, inner sep=0pt,
             label=below:{$S_3$},
             fit=(a3)] {};   
\node (af4)  [draw, inner sep=0pt,
             label=below:{$S_4$},
             fit=(a4)] {}; 

\path[->,-latex](a1) edge[bend left=30] node[pos=0.5,above] {$P(A,B)$} (a2);
\path[->,-latex](a2) edge[bend left=30] node[pos=0.5,above] {$P(B,A)$} (a3);
\path[->,-latex](a3) edge[bend left=30] node[pos=0.5,above] {$P(A,B)$} (a4);

\end{tikzpicture}

\vspace{.2cm}

\begin{tikzpicture}
\tikzset{odd/.style={draw,rectangle,fill=gray!20}}
\tikzset{even/.style={draw,rectangle}}

\node[odd] (b1) {A};
\node[even, right=1cm of b1] (b2) {B};
\node[even, right=0mm of b2] (b3) {B};
\node[odd, right=1cm of b3] (b4) {A};
\node[odd, right=0mm of b4] (b5) {A};
\node[even, right=1cm of b5] (b6) {B};

\node (bf1)  [draw, inner sep=0pt,
             label=below:{$Z_1$},
             fit=(b1)] {};
\node (bf2)  [draw, inner sep=0pt,
             label=below:{$Z_2,Z_3$},
             fit=(b2) (b3)] {};
\node (bf3)  [draw, inner sep=0pt,
             label=below:{$Z_4,Z_5$},
             fit=(b4) (b5)] {};   

\node (bf4)  [draw, inner sep=0pt,
             label=below:{$Z_6$},
             fit=(b6)] {};   

\path[->,-latex](b1) edge[bend left=30] node[pos=0.5,above] {$Q(A,B,2)$} (b2);
\path[->,-latex](b3) edge[bend left=30] node[pos=0.5,above] {$Q(B,A,2)$} (b4);
\path[->,-latex](b5) edge[bend left=30] node[pos=0.5,above] {$Q(A,B,1)$} (b6);

\end{tikzpicture}

\vspace{0.2cm}

\begin{tikzpicture}
\tikzset{odd/.style={draw,rectangle,fill=gray!20}}
\tikzset{even/.style={draw,rectangle}}

\node[odd] (c1) {1.0};
\node[even, right=1cm of c1] (c2) {-0.9};
\node[even, right=0mm of c2] (c3) {-1.1};
\node[odd, right=1cm of c3] (c4) {0.95};
\node[odd, right=0mm of c4] (c5) {1.2};
\node[even, right=1cm of c5] (c6) {-0.8};

\node (cf1)  [draw, inner sep=0pt,
             label=below:{$Y_1$},
             fit=(c1)] {};
\node (cf2)  [draw, inner sep=0pt,
             label=below:{$Y_2,Y_3$},
             fit=(c2) (c3)] {};
\node (cf3)  [draw, inner sep=0pt,
             label=below:{$Y_4,Y_5$},
             fit=(c4) (c5)] {};   

\node (cf4)  [draw, inner sep=0pt,
             label=below:{$Y_6$},
             fit=(c6)] {};   

\path[->,-latex](c1) edge[bend left=30] node[pos=0.5,above] {$\gamma_2^3(A,B)$} (c2);
\path[->,-latex](c3) edge[bend left=30] node[pos=0.5,above] {$\gamma_4^5(B,A)$} (c4);
\path[->,-latex](c5) edge[bend left=30] node[pos=0.5,above] {$\gamma_6^6(A,B)$} (c6);
\end{tikzpicture}

\caption{Example realisations of Markov chain $S_1^4$, semi-Markov chain $Z_1^6$, and hidden semi-Markov chain $Y_1^6$. Each block holds a realised value of the random variable below it, and the transitions define the probability of each realisation.}
\label{fig:smm}

\end{figure}

Our choice of notation is heavily inspired by \cite{Girardin2018, Barbu2008}, which gives a more detailed treatment of (hidden) semi-Markov processes. Refer to the concrete example in Fig. \ref{fig:smm}.

\section{Channel Model}
We now propose a channel model for the nanopore sequencer based on the concatenation of an i.i.d. duplication channel with a FSSMC. The channel will model random dwell times with i.i.d. duplications, ISI with a finite state-space, and noisy measurements with additive noise. The following makes the components of the illustration in Fig. \ref{fig:nnc} precise.

\vspace{1mm}
\textbf{Noiseless nanopore channel}.\quad Let $\Omega=\{A,T,C,G\}^\tau$ be the state-space of the nanopore sequencer with memory constraint $\tau\geq 1$. Given a sequence of bases $b_1^n$ in $\{A,T,C,G\}^n$, let us model the nanopore as a shift register that stores the subsequence $b_{\ell-\tau-1}^{\ell} = (b_{\ell-\tau-1},b_{\ell-\tau},\ldots, b_\ell)$ after $\ell$ bases have been shifted through, where the initial state is some $(b_{-\tau-1}, b_{-\tau},\ldots,b_0)\in\Omega$. In addition, let these subsequences induce a mean measurement level by injective mapping $f: \Omega \rightarrow \mathcal{Y}\subseteq \mathbb{R}$, as defined for the {\em noiseless nanopore channel} in  \cite{Hulett2021}. 

\vspace{1mm}
\textbf{Noisy nanopore channel}.\quad
Let $S_\ell = b_{\ell-\tau-1}^{\ell}$ be one of $4^\tau$ possible states in $\Omega$, then the subsequent state $S_{\ell+1}=b_{\ell-\tau}^{\ell+1}$ is only conditionally dependent on $S_{\ell}$. Hence, the sequence of states that are stored in the shift register form a Markov chain $(S_\ell)$ on $\Omega$. Now, let us say that we can observe the state of the shift register at {\em deterministic} time intervals to yield a sequence of {\em sample states} $(Z_1,Z_2,\ldots)$, however suppose that the bases are shifted in at {\em random} time intervals such that each state $S_\ell$ is observed for a {\em duration} of $K_\ell$ samples. In this scenario, we have that $Z_t = S_\ell$ for all $t$ such that $T_{\ell-1} = K_1 + \ldots + K_{\ell-1} < t \leq T_{\ell} = K_1 + \ldots + K_{\ell-1} + K_{\ell}$, where $T_\ell$ denotes the time when state $S_\ell$ is observed for the last time. Hence, the sequence of observed sample states in the shift register at different time instances form a semi-Markov chain $(Z_t)$ on $\Omega$. Finally, let us take {\em noisy measurements} $(Y_1,Y_2,\ldots)$ of the hidden sample states $(Z_1,Z_2,\ldots)$, where $Y_t = f(Z_t) + N_t$ for zero-mean additive noise $N_t$. This sequence of measurements form a hidden semi-Markov chain $(Y_t)$ on $\mathcal{Y}\subseteq \mathbb{R}$. 

We have now defined all components of the channel model, which are summarised in the following definition. 

\begin{definition}[Noisy nanopore channel]
Let us define $m$ channel-uses of the noisy nanopore channel as $W^{(m)} : (S_0,S_1,\ldots, S_m) \rightarrow (Y_1,\ldots,Y_{T_1}, Y_{T_1 + 1},\ldots,Y_{T_m})$, where the input $(S_0,S_1,\ldots, S_m)$ is a Markov chain on $\Omega$, and the output is derived by mapping the semi-Markov chain $(Z_1,\ldots,Z_{T_1},Z_{T_1+1},\ldots,Z_{T_m})$ on $\Omega$ to the hidden semi-Markov chain $(Y_1,Y_2,\ldots,Y_{T_m})$ on $\mathcal{Y}\subseteq \mathbb{R}$, where the dwell times $K_\ell = T_{\ell}-T_{\ell-1}$ are defined on $\Lambda \subseteq \mathbb{N} \backslash \{0\}$. The mapping of the output is the mean measurement level given by $f$ with zero-mean additive noise.\label{def:nnc}
\end{definition}

This natural definition of the noisy nanopore channel has hinted at a new class of channels, namely finite-state semi-Markov channels. If we set $\Lambda = \{1\}$, then the FSSMC collapses into the classical FSMC, and hence we say that the former generalises the latter. In addition, a FSSMC with geometrically distributed durations is also a FSMC (since the semi-Markov source collapses into a Markov source).

We briefly note that the zero element is removed from the durations $\mathsf{supp}(K) = \Lambda$ since that would imply concatenation of a deletion channel with deletion probability $d=\mathbb{P}(K=0)$, but we do not allow this here. Future work may study a generalisation of this channel, where deletion is allowed.

\section{Detection Algorithms}
We now generalise two important detection algorithms, namely the forward-backward algorithm and the Viterbi algorithm. In \cite{Yu2010, Girardin2018}, similar algorithms were derived for inference of the sample states $(Z_t)$ given the observations $(Y_t)$, but with an unknown number of states. In our case, we know $m$ channel-uses segments $(Z_t)$ into $m$ states of $S_1^m$, hence the detection problem is to find which non-overlapping segments of samples in $(Y_t)$ correspond to which sequence of states.

%Firstly, we derive the message passing algorithms to compute $\mathbb{P}(S_\ell|Y_1^{T_m})$, namely MAP symbol detection. Secondly, we derive a Viterbi algorithm that finds the most likely sequence of states $S_1^m$, namely MAP sequence detection.
%
\subsection{MAP symbol detection}
The goal of this section is to derive message passing algorithms to compute $\mathbb{P}(S_\ell| Y_1^{T_m})$ for all $\ell \in [m]$. The existing message passing algorithms \cite{Yu2010} for hidden semi-Markov chains compute $\mathbb{P}(Z_t| Y_1^{T_m})$ and $\mathbb{P}(Z_{t'}^t| Y_1^{T_m})$, however we are interested in the probability of the $\ell$-th segment being in a particular state given that there are exactly $m$ segments in total, which includes all sample paths $Z_1^{T_m}$ with $Z_{t'}^t = S_\ell$ for all $t',t \in [T_m]$ such that $t'<t$. Note that the constraint of $m$ segments effectively adds memory the size of the entire block.

\vspace{1mm}
\textbf{Generalised forward algorithm.}\quad Let us now generalise the classical forward algorithm. The initial forward probabilities are defined as
\begin{align}
    \alpha_{1, t}(s) &:=\mathbb{P}(Y_1^t, S_1 = s) \notag\\
    %&= \sum_{s' \in \Omega} \mathbb{P}(Y_{1}^{t}|Z_{1}^{t}=s) \mathbb{P}(K_\ell = t) P(s',s)\mu(s') \notag\\
    &= \sum_{s' \in \Omega} \gamma_1^t(s',s)\mu_0(s')
\end{align}
for all $t \in [T_m]$, and for all $s \in \Omega$. The inductive forward probabilities are defined as
\begin{align}
    \alpha_{\ell, t}(s) &:= \mathbb{P}(Y_1^t, S_\ell = s) \notag\\
    %&= \sum_{s' \in \Omega} \sum_{t' \in [t-1]} \mathbb{P}(Y_{t'+1}^{t}|Z_{t'+1}^{t}=s) Q(s',s',t-t') \notag\\
    &= \sum_{s' \in \Omega}\sum_{t' \in [t-1]} \gamma_{t'+1}^t(s',s) \alpha_{\ell-1, t'}(s')
\end{align}
for all $t \in [T_m]$, for all $\ell \in [m]\backslash\{1\}$, and for all $s \in \Omega$. Observe that setting $\Lambda = \{1\}$ yields the classical forward algorithm.

\vspace{1mm}
\textbf{Generalised backward algorithm.}\quad Let us now generalise the classical backward algorithm. The initial backward probabilities are defined to be $\beta_{m, T_m}(s) := 1$ for all $s \in \Omega$. The inductive backward probabilities are defined as
\begin{align}
    \begin{split}\beta_{\ell, t}(s) &:= \mathbb{P}(Y_{t+1}^{T_m}| S_\ell = s)\\
    %&= \sum_{s' \in \Omega}\sum_{t' \in [t+1:T]}\mathbb{P}(Y_{t+1}^{t'}|Z_{t+1}^{t'}=s') \\
    %& \quad\quad\quad\quad\quad\quad\quad\quad\quad\quad \cdot Q(s,s',t'-t) \beta_{t', \ell+1}(s')\\
    &= \sum_{s' \in \Omega}\sum_{t' \in [t+1:T_m]} \gamma_{t+1}^{t'}(s,s') \beta_{\ell+1, t'}(s')
    \end{split}
\end{align}
for all $t \in [T_m]$, for all $\ell \in [m-1]$, and for all $s \in \Omega$. Once again, setting $\Lambda = \{1\}$ yields the classical backward algorithm.

\vspace{1mm}
\textbf{Generalised forward-backward algorithm.}\quad
We now use the generalised forward algorithm and the generalised backward algorithm to compute the posterior probabilities
\begin{align}
    \psi_{\ell}(s) &:= \mathbb{P}(S_\ell = s| Y_1^{T_m}) \notag\\
   % &= \frac{\mathbb{P}(Y_1^T, S_\ell = s)}{\mathbb{P}(Y_1^T)}\notag\\
    &=  \frac{\sum_{t \in [T_m]} \mathbb{P}(Y_1^t, S_\ell = s) \mathbb{P}(Y_{t+1}^{T_m}| S_\ell = s) }{\mathbb{P}(Y_1^{T_m})}\notag\\
    &= \frac{\sum_{t \in [T_m]} \alpha_{ \ell, t}(s) \beta_{\ell, t}(s)  }{\sum_{s' \in \Omega} \alpha_{m, T_m}(s')}.
    \label{eq:fb1}
\end{align}
Similarly, we can also compute the joint posterior probabilities
\begin{align}
    \psi_{\ell}(s', s) &:= \mathbb{P}(S_{\ell-1} = s', S_\ell = s| Y_1^{T_m}) \notag\\
    %&= \frac{\mathbb{P}(Y_1^T, S_{\ell-1} = s', S_\ell = s)}{\mathbb{P}(Y_1^T)}\notag\\
    &= \frac{\sum_{t',t \in [T_m] : t' < t} \alpha_{\ell-1, t'}(s') \gamma_{t'+1}^{t}(s',s) \beta_{\ell, t}(s) }{\sum_{s'' \in \Omega} \alpha_{m, T_m}(s'')}.
    \label{eq:fb2}
\end{align}
Aside from detection, we will use both of these probabilities to estimate achievable rates of the NNC in Section V.

\subsection{MAP sequence detection}
The goal of this section is to derive a Viterbi algorithm to compute the sequence of states that maximise the posterior $\mathbb{P}(Z_1^{T_m}|Y_1^{T_m}) \propto \mathbb{P}(Z_1^{T_m}, Y_1^{T_m})$. Compared to the classical Viterbi algorithm, we not only sequentially maximise the path of states $S_1^m$, but also the sequence of jump times $T_1^{m}$. Importantly, we require that the first state begins at index $T_0+1$ ($T_0=0$) and the $m$-th state ends at index $T_m$ known at the decoder, hence we must segment an observation $Y_1^{T_m}$ into a sequence of $m$ segments that maximises the posterior.

\vspace{1mm}
\textbf{Generalised Viterbi algorithm.}\quad Let us now derive the dynamic program of the generalised Viterbi algorithm as
\begin{align}
    V(m,t_m) &:= \max_{\substack{s_1,\ldots,s_m \\ 1<t_1<\ldots < t_m}} \prod_{\ell=1}^{m} \gamma_{t_{\ell-1}+1}^{t_\ell}(s_{\ell-1},s_\ell) \label{eq:viterbi_dp}\\
    \begin{split}&= \max_{\substack{s_m \\ t_{m-1} < t_m}} \gamma_{t_{m-1}+1}^{t_m}(s_{m-1},s_m) \\ &\quad\quad\quad\cdot \left[\max_{\substack{s_1,\ldots,s_{m-1} \\ 1<t_1<\ldots<t_{m-1}}} \prod_{\ell=1}^{m-1} \gamma_{t_{\ell-1}+1}^{t_\ell}(s_{\ell-1},s_\ell)\right]\end{split}\notag\\
    &= \max_{\substack{s_m \\ t_{m-1} < t_m}} \gamma_{t_{m-1}+1}^{t_m}(s_{m-1},s_m) V(m-1,t_{m-1})\notag
\end{align}
where $V(0,0):=1$, noting that each $t_\ell$ is a constant in $V(\ell, t_\ell)$. The dynamic program of subproblem $V(\ell, t_\ell)$ jointly maximises the posterior of the sequence of states $s_1^\ell$ and the sequence of segment jump times $t_1^{\ell-1}$ such that the $\ell$-th segment jumps after time $t$. Since $V(\ell, t)$ depends on other subproblems $V(\ell', t')$ for all $\ell' < \ell$ and for all $t'<t$, it can be shown with some work that standard dynamic programming techniques solve (\ref{eq:viterbi_dp}) in a worst-case $O(m T_m 4 |\Omega| |\Lambda|)$ time (note that for the nanopore any state is connected to a maximum of four other states).

%A segmentation algorithm like this was studied in \cite{Han2004}, however the states were assumed to be i.i.d. In \cite{Jackson2005}, this class of algorithms was studied in a more general sense, where instead of a posterior they considered a ``fitness function''. While non-i.i.d. segments were not considered, the techniques surrounding this class of algorithms could be useful for fast implementations of (\ref{eq:viterbi_dp}).

\section{Achievable Rates}
We now turn our attention to computing achievable rates.
%, which are important for evaluating the performance of coding schemes that will be developed for this channel in the future.
The ultimate goal is to compute the capacity of Definition \ref{def:nnc}, constrained to Markov sources, which is given by

\begin{align}
\begin{split}
  C(W) &= \sup_{P} I(\mathscr{S}; \mathscr{Y})\\
  &= \lim_{m\rightarrow\infty}\frac{1}{m}\sup_{P} I(S_1^m; Y_1^{T_m})
  \label{eq:cap_def}
  \end{split}
\end{align}

in bits per base (or bits per state), where $P$ is a Markov kernel of Markov source $(S_\ell)$ and the resulting achievable rate is $I(\mathscr{S}; \mathscr{Y})$. As in the previous section, let us take methods for the FSMC and show analogous methods for the NNC.

In \cite{Kavcic2001}, the classical Blahut-Arimoto algorithm for discrete memoryless channels was generalised to compute achievable rates of finite-state channels. It was not rigorously proven to converge to the capacity, however it was strongly suggested based on their numerical results, and later proven to at least converge to a critical point \cite{Vontobel2008}. In this spirit, we show the (almost) direct application of GBAA to the NNC.

 GBAA iteratively searches for the Markov kernel that achieves the maximisation of (\ref{eq:cap_def}), computing achievable rates at each iteration using the method we now describe. Let $\mu$ be the stationary distribution of some Markov kernel $P$, then we can write the achievable information rate as
\begin{multline}
    \frac{1}{m}I(S_1^m; Y_1^{T_m}) =\\ \underbrace{\frac{1}{m}\sum_{\ell=1}^{m} H(S_\ell|S_{\ell-1})}_{\sum_{i,j \in \Omega} \mu(i) P(i,j) \log \frac{1}{P(i,j)}}-\frac{1}{m}\sum_{\ell=1}^{m} H(S_\ell|S_{\ell-1},Y_1^{T_m}).
    \label{eq:mi_rate}
\end{multline}
The first term has a closed-form, however the second term is significantly more challenging to compute. We decompose the second term identically to \cite{Kavcic2001}, which defines the so-called {\em $T$-values} as

\begin{align}
    T(i,j) &= \lim_{m\rightarrow\infty} \frac{1}{m}\sum_{\ell=1}^{m}  \mathbb{E}\left[\log \frac{\psi_{\ell}(i,j)^{\frac{\psi_{\ell}(i,j)}{\mu(i) P(i,j)}}}{ \psi_{\ell-1}(i)^{\frac{\psi_{\ell-1}(i)}{\mu(i)}} }\right]
    \label{eq:t_values},
\end{align}

 then, assuming we can compute these $T$-values, the achievable information rate of the NNC is computed as

\begin{align}
    I(\mathscr{S}; \mathscr{Y}) = \sum_{i,j \in \Omega} \mu(i) P(i,j) \log \left[\frac{1}{P(i,j)} + T(i,j)\right]
\end{align}

and may be optimised using GBAA \cite{Kavcic2001}. Similarly to \cite{Kavcic2001, Vontobel2008}, we conjecture that this algorithm converges to the capacity since (ergodic) finite-state semi-Markov processes are a subclass of (ergodic) finite-state processes (i.e., processes with memory). Strictly speaking, the capacity of a FSSMC must optimise semi-Markov kernel $Q$, but since $\mathbb{P}(K)$ is fixed by the duplication channel we need only consider Markov kernel $P$, where the information contribution of $K$ is subtracted to give (\ref{eq:mi_rate}) for the NNC and is identical in form to the FSMC.
\newpage
The $T$-values can be estimated using a variation of the Arnold-Loeliger method \cite{Arnold2001}, which assumes ergodicity to compute estimates of (\ref{eq:t_values}) by applying the classical forward-backward algorithm on a long realisation $y_1^{t_m}$. For the NNC, we require the probabilities $\psi_{\ell}(i,j)$ and $\psi_{\ell-1}(i)$, which  are (\ref{eq:fb2}) and (\ref{eq:fb1}), respectively, from Section IV.

\section{Numerical Results}
%\subsection{Example channel}
{\bf Model parameters.} Let us now consider which parameters of the NNC accurately model the actual nanopore channel. Scrappie \cite{Oxford2013} is a simulator by Oxford Nanopore Technologies that takes input bases $b_1^n$ and outputs tuples $\{(x_\ell,\sigma^2_\ell)\}_{\ell=1}^{n-\tau+1}$ for the mean and variance of measurements for the $\ell$-th base shifted into the nanopore. The model in Scrappie assumes the memory constraint $\tau$ is large, but we choose $\tau=5$ since it gives roughly stationary parameters. We generate the channel mapping $f$ by inputting multiple {\em de Bruijn sequences} (short sequences that enumerate all $4^\tau$ nucleotides) into Scrappie and averaging $x_\ell$ over identical states in each sequence. For simplicity, we use a constant $\sigma^2$ for all states, but all methods described earlier are easily extended. Noisy measurements are distributed as i.i.d.  $\mathsf{Normal}\bigl(f(s), \sigma^2\bigr)$ while in state $s\in\Omega$, and durations are distributed as i.i.d. $\mathsf{Uniform}(\Lambda)$.

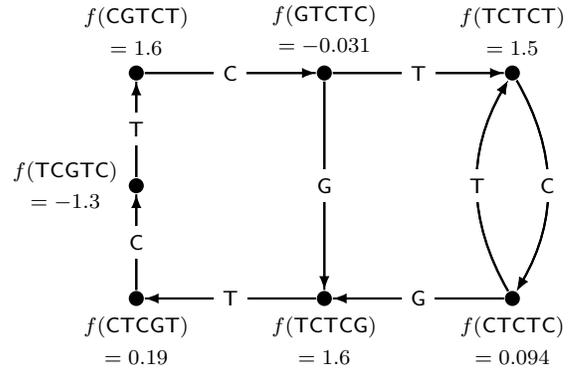
\begin{figure}[h!]
    \centering
    
    \begin{tikzpicture}[
      thick,
      acteur/.style={
        circle,
        fill=black,
        thick,
        inner sep=2pt,
        minimum size=0.2cm
      }
    ] 
      \node (a1) at (0,0) [acteur,label={[align=center]below:{\footnotesize $f(\mathsf{CTCGT})$\\\footnotesize$=0.19$}}]{};
      \node (a2) at (0,1.5)[acteur,label={[align=center]left:{\footnotesize $f(\mathsf{TCGTC})$\\\footnotesize$=-1.3$}}]{}; 
      \node (a3) at (0,3)
      [acteur,label={[align=center]:{\footnotesize $f(\mathsf{CGTCT})$\\\footnotesize$=1.6$}}]{}; 
      
      \node (a4) at (2.5,0) [acteur,label={[align=center]below:{\footnotesize $f(\mathsf{TCTCG})$\\\footnotesize$=1.6$}}]{}; 
      \node (a5) at (2.5,3) [acteur,label={[align=center]:{\footnotesize $f(\mathsf{GTCTC})$\\\footnotesize$=-0.031$}}]{}; 
      
      \node (a6) at (5,0) [acteur,label={[align=center]below:{\footnotesize $f(\mathsf{CTCTC})$\\\footnotesize$=0.094$}}]{};
      \node (a7) at (5,3) [acteur,label={[align=center]:{\footnotesize $f(\mathsf{TCTCT})$\\\footnotesize$=1.5$}}]{};

      \draw[->,-latex] (a1) -- (a2); 
      \draw (a1) -- (a2) node [midway, fill=white] {\footnotesize $\mathsf{C}$};

      \draw[->,-latex] (a2) -- (a3); 
      \draw (a2) -- (a3) node [midway, fill=white] {\footnotesize $\mathsf{T}$};
      
      \draw[->,-latex] (a3) -- (a5);
      \draw (a3) -- (a5) node [midway, fill=white] {\footnotesize $\mathsf{C}$};
      
      \draw[->,-latex,label={$c$}] (a5) -- (a7);
      \draw (a5) -- (a7) node [midway, fill=white] {\footnotesize $\mathsf{T}$};
      
      \path[->,-latex](a7) edge[bend left=30] node[pos=0.5,right] {} (a6);
      \path[->,-latex](a7) edge[bend left=30] node[pos=0.5,midway,fill=white] {\footnotesize $\mathsf{C}$} (a6);
      
      \path[->,-latex](a6) edge[bend left=30] node[pos=0.5,right] {} (a7);
      \path[->,-latex](a6) edge[bend left=30] node[pos=0.5,midway,fill=white] {\footnotesize $\mathsf{T}$} (a7);
      
      \draw[->,-latex] (a6) -- (a4);
      \draw (a6) -- (a4) node [midway, fill=white] {\footnotesize $\mathsf{G}$};
      
      \draw[->,-latex] (a4) -- (a1);
      \draw (a4) -- (a1) node [midway, fill=white] {\footnotesize $\mathsf{T}$};
      
      \draw[->,-latex] (a5) -- (a4);
      \draw (a4) -- (a5) node [midway, fill=white] {\footnotesize $\mathsf{G}$};
     
    \end{tikzpicture} 
    
    \caption{A jump-constrained graph derived from measurement levels in Scrappie \cite{Oxford2013}. The edges are channel inputs and the nodes are a state $s\in\Omega$ with channel mapping $f(s)$.}
    \label{fig:ssplot}
\end{figure}

Although $P$ is sparse for $\tau=5$, it yields a state-space of $4^5 = 1024$ states that is prohibitive for the algorithms of Section IV. Hence, we reduce the state-space by removing edges in the graph of the Markov source that are not desirable for synchronisation (segmentation), namely the edges that have a {\em jumping distance}, $\abs{f(i)-f(j)}$, below a minimum threshold $J_{min}$, for all $i,j\in \Omega$ with $P(i,j)>0$. Ensuring the new graph is {\em irreducible}, and thus ergodic, we use standard algorithms to compute the list of {\em strongly connected components} and take the one with largest entropy. Set $J_{min} = 1.38$ to get our example state-space graph in Fig. \ref{fig:ssplot}. 
\vspace{1mm}

\begin{figure}[h]
\vspace{-2mm}
\centering
\begin{tikzpicture}[scale=0.8]

\begin{axis}[%
at={(0.758in,0.481in)},
scale only axis,
xmin=0.1,
xmax=0.4,
xlabel style={font=\color{white!15!black}},
xlabel={$\sigma$},
ymin=0.05,
ymax=0.34,
ylabel style={font=\color{white!15!black}},
ylabel={Rate (bits/base)},
axis background/.style={fill=white},
title style={font=\bfseries},
title={i.u.d. jump-constrained achievable information rates},
xmajorgrids,
ymajorgrids,
legend style={at={(0.03,0.03)}, anchor=south west, legend cell align=left, align=left, draw=white!15!black}
]
\addplot [color=white!75!gray, line width=1.0pt, mark size=4.0pt, mark=o, mark options={solid, white!75!gray}]
  table[row sep=crcr]{%
0.1	0.3064\\
0.15	0.3064\\
0.2	0.3064\\
0.25	0.3065\\
0.3	0.3064\\
0.35	0.3066\\
0.4	0.3066\\
};
\addlegendentry{$\Lambda=\{1\}$}

\addplot [color=lightgray, line width=1.0pt, mark size=2.7pt, mark=triangle, mark options={solid, lightgray}]
  table[row sep=crcr]{%
0.15	0.3062\\
0.2	0.3063\\
0.25	0.3062\\
0.3	0.3063\\
0.35	0.3062\\
0.4	0.3059\\
};
\addlegendentry{$\Lambda=\{2,3\}$}

\addplot [color=white!50!darkgray, line width=1.0pt, mark size=2.7pt, mark=triangle, mark options={solid, rotate=270, white!50!darkgray}]
  table[row sep=crcr]{%
0.15	0.3062\\
0.2	0.3062\\
0.25	0.3063\\
0.3	0.3059\\
0.35	0.2428\\
0.4	0.2478\\
0.45	0.192\\
0.5	0.1238\\
};
\addlegendentry{$\Lambda=\{2,3,\ldots,6\}$}

\addplot [color=gray, line width=1.0pt, mark size=5pt, mark=diamond, mark options={solid, gray}]
  table[row sep=crcr]{%
0.1	0.3062\\
0.15	0.3063\\
0.2	0.3063\\
0.25	0.3005\\
0.3	0.2419\\
0.35	0.1694\\
0.4	0.1349\\
};
\addlegendentry{$\Lambda=\{1,2\}$}

\addplot [color=black!25!gray, line width=1.0pt, mark size=2.8pt, mark=square, mark options={solid, black!25!gray}]
  table[row sep=crcr]{%
0.1	0.3062\\
0.15	0.3063\\
0.2	0.3044\\
0.25	0.2531\\
0.3	0.1634\\
0.35	0.1081\\
0.4	0.086\\
};
\addlegendentry{$\Lambda=\{1,2, 3\}$}

\addplot [color=darkgray, line width=1.0pt, mark size=2.7pt, mark=triangle, mark options={solid, rotate=180, darkgray}]
  table[row sep=crcr]{%
0.1	0.3062\\
0.15	0.3054\\
0.2	0.3061\\
0.25	0.2047\\
0.3	0.1307\\
0.35	0.0985\\
0.4	0.0744\\
};
\addlegendentry{$\Lambda=\{1,2,\ldots,5\}$}

\addplot [color=black!50!darkgray, line width=1.0pt, mark size=4.0pt, mark=x, mark options={solid, black!50!darkgray}]
  table[row sep=crcr]{%
0.1	0.3062\\
0.15	0.3063\\
0.2	0.3036\\
0.25	0.2051\\
0.3	0.1393\\
0.35	0.0977\\
0.4	0.0591\\
};
\addlegendentry{$\Lambda=\{1,2,\ldots,10\}$}

\node[above] at (50,260) {$H_{max}=0.3063$};

\end{axis}
\end{tikzpicture}%
\caption{Achievable rates using the independent and uniformly distributed (i.u.d.) input distribution for the example in Fig. \ref{fig:ssplot}.}
\label{fig:info_rates}
\vspace{-4mm}
\end{figure}
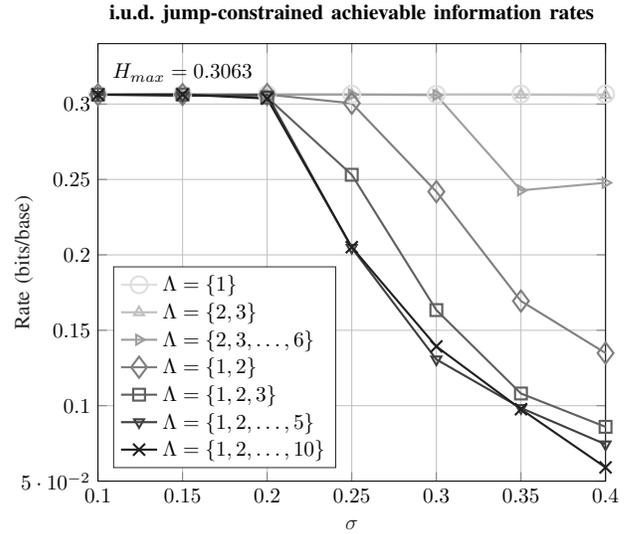

{\bf Achievable rates.} We now discuss the achievable information rates in Fig. \ref{fig:info_rates}, which are plotted over different values of $\sigma$ and $\Lambda$ to highlight the effect of the minimum duration and the range of durations with respect to measurement noise. The achievable rates are estimated using a limited $m=10,000$ channel-uses. We choose $\sigma$ to be in the range $0.10-0.40$, which covers those given by the Scrappie simulator. 

We observe the rates with durations $\Lambda=\{1\}$, i.e., the NNC without duplications, which achieve the maximum source entropy $H_{\max}=0.3063$ over this range of $\sigma$. We further observe that  single duplications with $\Lambda = \{1,2\}$ significantly degrade the rates, but fortunately this degradation halts as the range is widened to $\Lambda=\{1,2,\ldots,5\}$, giving identical rates to $\Lambda=\{1,2,\ldots,10\}$ and beyond. Additionally, we observe that shifting the durations by one significantly improves the rate, which is demonstrated from $\Lambda = \{1,2\}$ to $\Lambda=\{2,3\}$, and from $\Lambda=\{1,2,\ldots,5\}$ to $\Lambda=\{2,3,\ldots,6\}$. Note the range effects remain. This observation confirms our intuition that the shortest duration is the most difficult to synchronise in a noisy channel, especially in the high-noise regime.

\vspace{3.5mm}
\section{Conclusion}
In summary, we proposed the NNC for modelling the dominant distortions of the nanopore sequencer. We filled the gap in previous works by including duplications in the model, while maintaining tractability. The tractability of the model was demonstrated by the detection algorithms that generalised their classical counterparts. We estimated achievable rates for an example NNC using methods closely tied to GBAA.

This paper established a model with basic tools before coding schemes are considered in our future work. Such coding schemes will have to satisfy the source constraints required for synthesis and sequencing of DNA, all while being efficiently decodable. Our detection algorithms open the possibility of iterative decoding of LDPC codes with the generalised forward-backward algorithm, or sequential decoding of convolutional codes with the generalised Viterbi algorithm. We are hopeful that this avenue of research will lead to reliable nanopore sequencing.

\clearpage
\newpage
\IEEEtriggeratref{8}
\bibliography{refs}

\begin{thebibliography}{10}
\providecommand{\url}[1]{#1}
\csname url@samestyle\endcsname
\providecommand{\newblock}{\relax}
\providecommand{\bibinfo}[2]{#2}
\providecommand{\BIBentrySTDinterwordspacing}{\spaceskip=0pt\relax}
\providecommand{\BIBentryALTinterwordstretchfactor}{4}
\providecommand{\BIBentryALTinterwordspacing}{\spaceskip=\fontdimen2\font plus
\BIBentryALTinterwordstretchfactor\fontdimen3\font minus
  \fontdimen4\font\relax}
\providecommand{\BIBforeignlanguage}[2]{{%
\expandafter\ifx\csname l@#1\endcsname\relax
\typeout{** WARNING: IEEEtran.bst: No hyphenation pattern has been}%
\typeout{** loaded for the language `#1'. Using the pattern for}%
\typeout{** the default language instead.}%
\else
\language=\csname l@#1\endcsname
\fi
#2}}
\providecommand{\BIBdecl}{\relax}
\BIBdecl

\bibitem{ONT}
{O}xford~{N}anopore {T}echnologies, ``Nanopore sequencer,''
  https://nanoporetech.com.

\bibitem{Lopez2019}
R.~Lopez, Y.-J. Chen, S.~D. Ang, S.~Yekhanin, K.~Makarychev, M.~Z. R{\'a}cz,
  G.~Seelig, K.~Strauss, and L.~Ceze, ``{DNA} assembly for nanopore data
  storage readout,'' \emph{Nature Communications}, vol.~10, 2019.

\bibitem{Laszlo2014}
A.~H. Laszlo, I.~Derrington, B.~C. Ross, H.~Brinkerhoff, A.~Adey, I.~C. Nova,
  J.~M. Craig, K.~W. Langford, J.~Samson, R.~Daza, K.~Doering, J.~Shendure, and
  J.~Gundlach, ``Decoding long nanopore sequencing reads of natural {DNA},''
  \emph{Nature biotechnology}, vol.~32, pp. 829 -- 833, 2014.

\bibitem{Mao2017}
W.~Mao, S.~Diggavi, and S.~Kannan, ``Models and information-theoretic bounds
  for nanopore sequencing,'' \emph{2017 IEEE International Symposium on
  Information Theory (ISIT)}, pp. 2458--2462, 2017.

\bibitem{Hulett2021}
R.~Hulett, S.~Chandak, and M.~Wootters, ``On coding for an abstracted nanopore
  channel for {DNA} storage,'' \emph{2021 IEEE International Symposium on
  Information Theory (ISIT)}, pp. 2465--2470, 2021.

\bibitem{Tang2019}
Y.~Tang and F.~Farnoud, ``Error-correcting codes for noisy duplication
  channels,'' \emph{2019 57th Annual Allerton Conference on Communication,
  Control, and Computing (Allerton)}, pp. 140--146, 2019.

\bibitem{Conde2018}
L.~Conde-Canencia and L.~Dolecek, ``Nanopore {DNA} sequencing channel
  modeling,'' in \emph{2018 IEEE International Workshop on Signal Processing
  Systems (SiPS)}, 2018, pp. 258--262.

\bibitem{Gallager1968}
R.~G. Gallager, ``Information theory and reliable communication,'' 1968.

\bibitem{Kavcic2001}
A.~Kavcic, ``On the capacity of {M}arkov sources over noisy channels,''
  \emph{GLOBECOM'01. IEEE Global Telecommunications Conference (Cat.
  No.01CH37270)}, vol.~5, pp. 2997--3001 vol.5, 2001.

\bibitem{Arnold2001}
D.-M. Arnold and H.-A. Loeliger, ``On the information rate of binary-input
  channels with memory,'' \emph{ICC 2001. IEEE International Conference on
  Communications. Conference Record (Cat. No.01CH37240)}, vol.~9, pp.
  2692--2695 vol.9, 2001.

\bibitem{Vontobel2008}
P.~O. Vontobel, A.~Kavcic, D.-M. Arnold, and H.-A. Loeliger, ``A generalization
  of the {B}lahut–{A}rimoto algorithm to finite-state channels,'' \emph{IEEE
  Transactions on Information Theory}, vol.~54, pp. 1887--1918, 2008.

\bibitem{Oxford2013}
{O}xford~{N}anopore {T}echnologies, ``Scrappie: a technology demonstrator for
  the {O}xford {N}anopore {R}esearch {A}lgorithms group,''
  https://github.com/nanoporetech/scrappie, 2017.

\bibitem{Barbu2008}
V.~S. Barbu and N.~Limnios, ``Semi-{M}arkov chains and hidden semi-{M}arkov
  models toward applications: Their use in reliability and {DNA} analysis,''
  2008.

\bibitem{Girardin2018}
V.~Girardin and N.~Limnios, ``{M}arkov and semi-{M}arkov processes,'' 2018.

\bibitem{Yu2010}
S.~Yu, ``Hidden semi-{M}arkov models,'' \emph{Artif. Intell.}, vol. 174, pp.
  215--243, 2010.

\end{thebibliography}
\bibliographystyle{IEEEtran}

\end{document}